\documentclass[aps,pra,nofootinbib,superscriptaddress,showpacs,twocolumn,longbibliography]{revtex4-2}
\usepackage{graphicx,epstopdf}
\usepackage{multirow}
\usepackage{mathtools}
\usepackage{amsmath}
\usepackage{amssymb,mathptmx}
\usepackage{fouridx}
\usepackage{graphicx}
\usepackage{amsthm}
\usepackage{eucal}
\usepackage{bm}
\usepackage{upgreek}
\usepackage{framed}
\usepackage{mathrsfs}
\usepackage{latexsym}
\usepackage{float}
\usepackage{subfigure}
\usepackage{color}
\usepackage{graphicx} 
\usepackage{bm,hyperref,tikz}
\usepackage[utf8]{inputenc} 
\usepackage{bbold}
\usepackage{bbm}
\usetikzlibrary{decorations.pathmorphing, patterns,shapes}

\newcommand{\be}{\begin{equation}}
	\newcommand{\ee}{\end{equation}}
\newcommand{\bea}{\begin{eqnarray}}
	\newcommand{\eea}{\end{eqnarray}}
\newcommand{\ba}{\begin{array}}
	\newcommand{\ea}{\end{array}}

\newcommand{\ket}[1]{\left\vert#1\right\rangle}

\newcommand{\Ba}{\mathrm{B}_\alpha\,}

\newcommand{\E}[1]{\mathrm{e}^{\mbox{\footnotesize$#1$}}}

\begin{document}	
	
\title{Highly photon loss tolerant quantum computing using hybrid qubits}

\author{S. Omkar}
\email{omkar.shrm@gmail.com}
\affiliation{Department of Physics and Astronomy, Seoul National University, 08826 Seoul, Korea}
\author{Y. S. Teo}
\affiliation{Department of Physics and Astronomy, Seoul National University, 08826 Seoul, Korea}
\author{Seung-Woo Lee}\affiliation{Center for Quantum Information, Korea Institute of Science and Technology, Seoul, 02792, Korea}
\author{H. Jeong}
\email{h.jeong37@gmail.com}
\affiliation{Department of Physics and Astronomy, Seoul National University, 08826 Seoul, Korea}

\begin{abstract}
We investigate a scheme for topological quantum computing using optical hybrid qubits and make an extensive comparison with previous all-optical schemes.  We show that the photon loss threshold reported by Omkar {\it et al}. [Phys. Rev. Lett. 125, 060501 (2020)] can be improved further by employing postselection and multi-Bell-state-measurement based entangling operation to create a special cluster state, known as Raussendorf lattice for topological quantum computation. In particular, the photon loss threshold is enhanced up to $5.7\times10^{-3}$, which is the highest reported value given a reasonable error model.
This improvement is obtained at the price of consuming more resources by an order of magnitude, compared to the scheme in the aforementioned reference. Neverthless, this scheme remains resource-efficient compared to other known optical schemes for fault-tolerant  quantum computation.


\end{abstract}
	
\maketitle

\section{Introduction}

The quantum optical platforms have not only the advantage of supplying quicker gate operations compared to the decoherence time~\cite{RP10} but also relatively efficient readouts, which makes them suitable platforms and one of the strongest contenders for realizing scalable quantum computation (QC). However, in these platforms, photon loss is ubiquitous which leads to optical qubit loss and is also a major source of noise, i.e., dephasing or depolarizing~\cite{RP10}, also known as the computational errors. 
Noise stands as the major obstacle in the path towards scalable QC. To overcome the effects of noise, we need fault-tolerant schemes that employ quantum error correction (QEC)~\cite{NC10, LD13}. QEC promises the possibility to realize a scalable QC with faulty qubits, gates and readouts (measurements), provided the noise level is below certain {\it threshold}. This threshold value is determined according to the details of the fault-tolerant (FT) architecture and the associated noise model. Moreover, QEC has also been employed  in quantum metrology~\cite{SMJL18,TOJ19} and communication~\cite{Ram14,Loock17,LRJ19}. References~\cite{OSB15,OSB15s,OSB16} showed that  QEC codes can also be used for efficiently characterizing a quantum dynamical maps that could be either completely positive or not~\cite{SL09, OSB15t}. 

Fault-tolerant schemes implemented with various kinds of optical qubits provide different ranges of tolerance
against both qubit loss and  computational errors. The parameters that determine the performance of a  fault-tolerant optical scheme are (i) photon loss and computational error thresholds and (ii) their operational values, (iii) {\it logical} error rate and (iv) resources incurred per logical gate operation. Logical error rate is the rate of failure of QEC that results in a residual error at the highest logical level of encoding~\cite{LD13,NC10}. 
From the {\it threshold theorem}~\cite{NC10, ND05}, we know that when the fault-tolerant optical hardware operates below the noise threshold,  the logical errors rate can be made arbitrarily close to zero by allocating more resources.  
Thus, operational values of the noise, i.e., photon loss and computational error rates too are important parameters as they determine the required resource to attain the target logical error rate.

It has recently  been demonstrated that by using  optical hybrid qubits entangled in the continuous-discrete optical domain, many shortcomings faced individually by continuous variable (CV) and discrete variable (DV) qubits can be overcome in linear optical quantum computing~\cite{LJ13,OTJ19}. 
In fact, the FTQC schemes based on either DV or CV qubits not only tend to have low thresholds and operational values for photon loss and computational error, but they also require extravagant resources to provide arbitrarily small logical error rates.
In oder to overcome these limitations, the scheme in Ref.~\cite{LJ13} uses optical hybrid qubits that combine  single-photon qubits \cite{KLM01} together with the coherent-state qubits 
\cite{JKL01,JK02,Ralph03,LRH08,MR11} that are a particular type of CV qubits with coherent states.
While this scheme offers an improvement in resource efficiency, both the threshold and operational values of the noise remain low as it employs CSS (Calbank-Shor-Steane) QEC codes~\cite{SteanPRL96,CS96,SeanPRSL96}.  
Our recent proposal for topological FTQC~\cite{OTJ19} employing special cluster states of optical hybrid qubits, also known as {\it Raussendorf} lattice ($\ket{\mathcal{C}_\mathcal{L}}$), exhibits an improvement in both operational and threshold values of photon loss and computational error  by an order of magnitude.  This hybrid-qubit-based topological FTQC (HTQC) scheme also offers the best resource efficiency.

HTQC uses  linear optics, optical hybrid states and Bell-state measurement (BSM) as entangling operation (EO) to create a  $\ket{\mathcal{C}_\mathcal{L}}$.  Interestingly, HTQC  does not involve postselection, and active switching is hence unnecessary. Furthermore, there is no need for in-line feed-forward operations. Therefore, HTQC is ballistic in nature. In this work we show that by employing postselection over the successful EOs  and using multi-BSM  EO  (see Fig.~\ref{fig:nHBSM}) at a certain stage of creation of $\ket{\mathcal{C}_\mathcal{L}}$, the photon-loss threshold, $\eta_{\rm th}$ can be further improved. We shall also show that this improvement costs more resources  than the HTQC, but  only by an order of magnitude.

The rest of the article is organized as follows. In Sec.~\ref{sec:mbtqc}, we briefly explain the preliminaries of measurement-based fault-tolerant topological QC  on a $\ket{\mathcal{C}_\mathcal{L}}$. Readers familiar with  the topic can skip this section. In Sec.~\ref{sec:phtqc} we describe our new scheme that employs postselection and multi-BSM-based EO to build $\ket{\mathcal{C}_\mathcal{L}}$ using hybrid qubits. Further, in Sec.~\ref{sec:star}, we detail the generation of star cluster states used as building-blocks for $\ket{\mathcal{C}_\mathcal{L}}$. In Sec.~\ref{sec:noise}, we describe the noise model used and simulation procedure of QEC is outlined in Sec.~\ref{sec:sim}. In Sec.~\ref{sec:result},  we present our results on the improved photon loss thresholds, and the details about resource estimation is provided in  the Sec.~\ref{sec:resource}. 
In Sec.~\ref{sec:compare}, we compare the various performance parameters of our scheme with those of other schemes for optical FTQC. 
Finally,  discussion and conclusion are presented in Sec.~\ref{sec:conclusion}.

\section{ Preliminaries}
\label{sec:mbtqc}
\begin{figure}[t]
\includegraphics[width=.9\columnwidth]{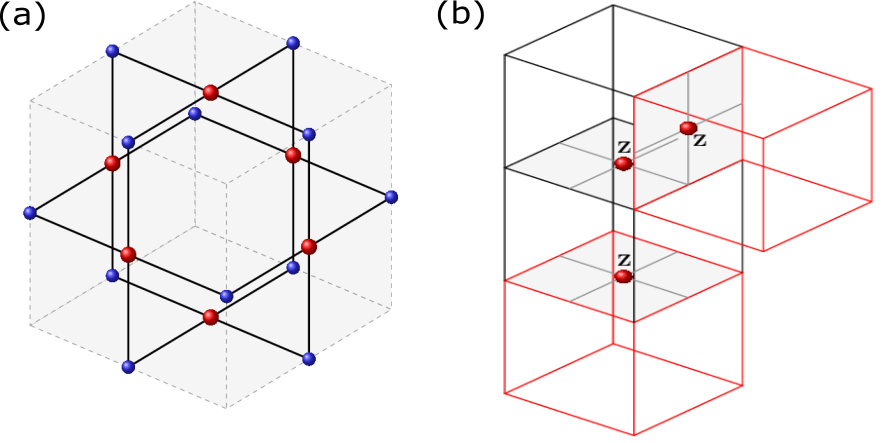}
\caption{(a) A unit-cell that makes up the lattice $\ket{\mathcal{C}_\mathcal{L}}$. The  qubits in red (larger) on the faces of the unit cell correspond to the primal-lattice and the others in blue (smaller) correspond to the dual-lattice.  The black (thick) lines  represent the presence of entanglement between the qubits. (b) A string of phase-flip errors will have detection events (red cells) only at the endpoints. 
}
\label{fig:cube}
\end{figure}

In this section we briefly review the measurement-based fault-tolerant topological QC on $\ket{\mathcal{C}_\mathcal{L}}$. For this purpose, we first define what the cluster states are in general and describe measurement-based FTQC on them. 
As an alternative to the circuit-based model for QC, Raussendorf and Briegel~\cite{RB01} developed a model where a universal set of gates can be realized using only adaptive single-qubit measurements in different bases on a multi-qubit entangled state known as  cluster state.  
In general, a cluster state, $\ket{\mathcal{C}}$ over a collection of qubits $\mathcal{C}$, is a state stabilized by the operators  $X_a\bigotimes_{b\in {\rm nh}(a)} Z_b$, where $a,b\in\mathcal{C}$, $Z_i$ and $X_i$ are the Pauli operators on the $i$th qubit, nh(a) denotes the adjacent neighborhood of qubit $a\in\mathcal{C}$. A multi-qubit $\ket{\mathcal{C}}$ has the form:  
\be
\ket{\mathcal{C}}=\prod_{b\in \rm{ nh}(a)}\textrm{CZ}_{a,b}\ket{+}_a\ket{+}_b,~\forall a\in \mathcal{C},
\label{eq:clus}
\ee
 where $\ket{\pm}=(\ket{0}\pm\ket{1})/\sqrt{2}$ is the eigenstate of $X$, while $\ket{0},\ket{1}$ are those of $Z$. ${\rm CZ}_{a,b}$, an EO, applies $Z$ on the target qubit $b$ if  the  source qubit $a$ is in the state $\ket{1}$. The {\it unit cell} shown in  Fig.~\ref{fig:cube}(a) is an example of a cluster state.
This measurement-based QC model is not fault-tolerant by nature and in order to achieve robustness against noise, the cluster-qubits were encoded into 5-qubit QEC codes~\cite{JF09} and Steane QEC codes~\cite{FY10}.

Another route to fault tolerance is to recognize that certain cluster states correspond to topological QEC codes. 
{\it Surface codes}, a class of topological QEC codes on  2D cluster states are known to provide a high error threshold of $\sim1\%$~\cite{WFH11} against computational errors. 
 It is known that surfaces codes can  tolerate neither qubit-loss nor EO failures; thus, are not suitable for optical platforms~\cite{o01, F12}.
The shortcomings of surface codes can be overcome by using a $\ket{\mathcal{C}_\mathcal{L}}$  for topological QEC. For a review on the topic refer to Refs.~\cite{FG09, WF14, BBD+09}. 
Topological QEC on $\ket{\mathcal{C}_\mathcal{L}}$~\cite{RHG06}
is known to provide a high error-threshold of 0.75\%~\cite{ RHG07, RH07} against computational errors that occur during preparation, storage, gate application and measurement. 
In addition, $\ket{\mathcal{C}_\mathcal{L}}$ can tolerate qubit-loss~\cite{BS10, WF14} and  {\it missing edges}~\cite{LDSB10} due to failed EOs, making it suitable for linear optical platforms.

\subsection{Error detection and correction}
\label{subsec:edc}
The lattice $\ket{\mathcal{C}_\mathcal{L}}$ can be thought of as a lattice formed by unit-cell aggangement as shown in the Fig.~\ref{fig:cube}(a).   This lattice has qubits  mounted on its faces and edges~\cite{RHG06}.   For QEC and QC , it is important to recognize that $\ket{\mathcal{C}_\mathcal{L}}$ is formed by inter-locking two types of lattices, namely the primal and dual lattices. The dual-lattices is a result of mapping the face-qubits of primal-lattice to edge-qubits and vice versa.  From Eq.~\eqref{eq:clus}, it is clear that each face of a unit-cell is stabilized by $X_i\bigotimes_{b} Z_b$ where $X_i$ denotes the $X$ operator on the $i$-th face of the unit-cell and $Z_b$ denotes the $Z$ operators on the boundary of the face.
A stabilizer of a unit cell associated with the primal-lattice is given by the product of six constituent face stabilizers i.e.,  $S_p=X_1X_2X_3X_4X_5X_6$.  To measure $S_p$, one needs to perform single-qubit measurements in the $X$-basis and multiply the individual outcomes.  When there is no phase-flip error on an odd number of qubits, the measurement outcome would be $s_p=+1$. 

\begin{figure}[t]
\includegraphics[width=.9\columnwidth]{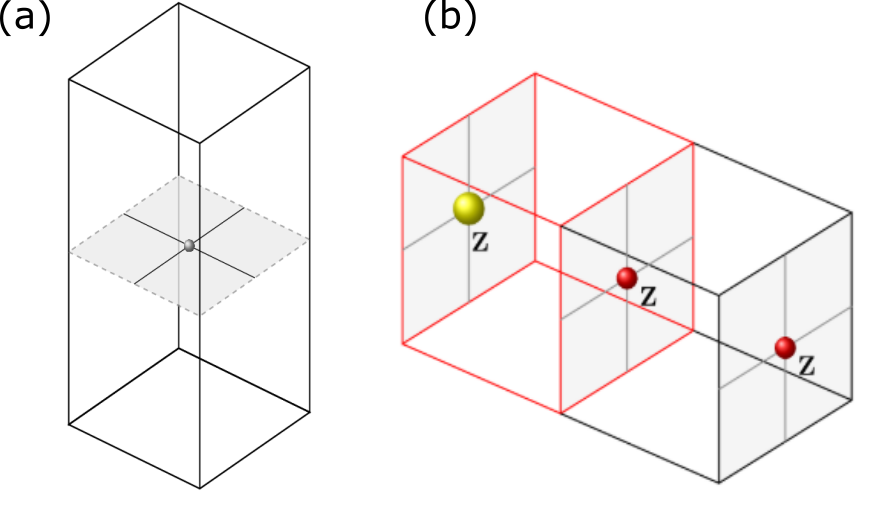}
\caption{(a)  The qubit on the common face of adjacent cells is considered to be lost. Stabilizers of the two adjacent cells can be multiplied to form a larger cell, which removes the dependency on the measurement outcome of the shared qubit. This feature is employed to deal with qubit-loss of unit-cells where the larger cell can perform error detection that is not possible by incomplete unit-cells.  
(b) Two unit-cells forming a distance $d=3$ codes is shown. Both errors (bigger yellow ball) on a single qubit  and two qubits (smaller red balls) cause the same detection events indicated by red cell.  As the single-error case has smaller weight, the MWPM always chooses it even if the errors occurred  on two qubits. When the error inference is wrong, making error correction by applying $Z$ on the larger qubit will complete the error chain connecting the two boundaries, causing a logical error.
}
\label{fig:logical}
\end{figure}

As the $Z$ operator on an odd number of face-qubits anti-commute with the $S_p$, the stabilizer measurement outcome would be $s_p=-1$. On the other hand, an even number of phase-flips go undetected as they commute with $S_p$ and have $s_p=+1$. Therefore, when $s_p=-1$, one can only detect but not locate the errors.  An error can be detected and located on $\ket{\mathcal{C}_\mathcal{L}}$ by measuring $S_p$ of the adjacent cubes as shown in Fig.~\ref{fig:cube}(b). Multiple errors on the adjacent cells form an  error-chain in  $\ket{\mathcal{C}_\mathcal{L}}$ that can be detected at its end points with the value $s_p=-1$ as shown in Fig.~\ref{fig:cube}(b). However, this only reveals the existence of an error chain, but does not locate every error.  Thus, one would need to guess the most likely error-chain and apply appropriate correction. This guess can be carried out using the efficient minimum weight perfect matching algorithm (MWPM)~\cite{Edmond65}. MWPM can make wrong guesses and may lead to {\it logical errors} discussed in the subsequent subsection. We note that the bit-flip errors on $\ket{\mathcal{C}_\mathcal{L}}$ have trivial effect and thus only the phase-flips are of concern in this QEC scheme.

In order to detect  errors on qubits other than those on the faces we invoke the concept of the dual-lattice where the edge-qubits in the primal-lattice are now the face-qubits. 
One can construct a unit-cell and stabilizer $S_d$ on the dual-lattice and carry out QEC just like the procedures on the primal-lattice. It is important to note that QEC on both type of  lattices proceeds independently.

\emph{Handling qubit losses:} When the qubits in the lattice are lost, it becomes impossible to measure the stabilizers $S_p$ or $S_d$ and detect the errors. To circumvent this issue, one can form a larger stabilizer by multiplying the two adjacent-cell stabilizers such that the lost qubit is shared between them. This eliminates the dependency of the stabilizer on the lost qubit as shown in Fig.~\ref{fig:logical}(a). The resultant  stabilizer with 10 $X$ operators can perform error detection just like a regular stabilizer of a unit-cell. If there are chain of losses, the same procedure can be extended to form larger cells that can replace unit-cells~\cite{BS10}.

\subsection{Logical operations and logical errors}
A few chosen qubits are measured in the $Z $basis to create defects that initialize the logical states on the  $\ket{\mathcal{C}_\mathcal{L}}$. This removes the qubits from the lattice and disentangles the qubits inside the measured region from the rest of the lattice. Depending on the chosen lattice type, the logical qubits would either be of primal or dual types.  Logical operations on the logical states correspond to a chain of $Z$ operators  that either  encircles a defect or connects two defects of the same type~\cite{RHG06,RHG07}. 
Equivalently, in the absence of defects, logical error happens when boundaries of same the type are connected by a chain of $Z$ operators.  

The code-distance $d$ is defined as the minimum number of $Z$ operations required to change the logical state of $\ket{\mathcal{C}_\mathcal{L}}$. Errors on the logical states can also be introduced  due to wrong inference  by the MWPM during QEC. An error chain of length $(d+1)/2$ or longer can lead to such wrong inferences. For example, consider two cells as shown in the Fig.~\ref{fig:logical}(b), forming a distance $d=3$ code where both the single-qubit error (bigger ball) and the two-qubit error (smaller balls) cause the same detection events.  As the single-error case has smaller weight, the MWPM preferencially chooses it even when errors have actually occurred  on the other two qubits. In this case, performing error correction by applying single $Z$ (on the larger qubit) will connect the two boundaries  causing a logical error.

\subsection{Universal gates}
Once a faulty $\ket{\mathcal{C}_\mathcal{L}}$ with missing qubits and phase-flip errors is available,  topological FTQC is carried out  by making sequential single-qubit measurements in  $X$ and $Z$ bases as dictated by the quantum algorithm being implemented. These defects are braided to achieve two-qubit logical operations topologically~\cite{RHG06,RHG07}. 
The lattice qubits are measured in  $X$ basis, the outcomes of which which not only provide error syndromes  but also effect  Clifford gates on the logical states. It is to be noted that not only tolerance against qubit losses but also two-qubit logical operations becomes available by moving from surface codes to 3D cluster-based QEC codes. 
The universal set of operations for QC is complete with inclusion of  {\it magic-state distillation} for which measurements on the chosen qubits are carried out  in the $(X\pm Y)/\sqrt{2}$ basis~\cite{RHG06,RHG07}.

\section{Raussendorf lattice with postselection and multi-BSM entangling operation}
\label{sec:phtqc}
In this work, similarly to HTQC,  a  $\ket{\mathcal{C}_\mathcal{L}}$ is created with optical hybrid qubits of the form: 
\be
\ket{\Psi_\alpha}=(\ket{\alpha}\ket{\textsc{h}}+\ket{-\alpha}\ket{\textsc{v}})/\sqrt{2},
\label{eq:raw}
\ee
where  $\ket{\textsc{h}}$, $\ket{\textsc{v}}$ are the discrete orthonormal polarization eigenkets of $Z$, and $\{\ket{\alpha}\ket{\textsc{h}},\ket{-\alpha}\ket{\textsc{v}}\}$ forms the computational basis for hybrid qubits where $\alpha$ is assumed to be real without loss of generality. 
However, here we add two extra features; postselection and the multi-BSM EO to improve $\eta_{\rm th}$ over HTQC. postselection will avoid the formation of undesired {\it diagonal} edges and thus resulting in a better $\ket{\mathcal{C}_\mathcal{L}}$. Employing the multi-BSM EOs will reduce the value of $\alpha$ required to build  $\ket{\mathcal{C}_\mathcal{L}}$. We demonstrate in  Sec.~\ref{sec:noise} that   using larger $\alpha$ invites larger dephasing on the hybrid qubits in the presence of photon loss. Therefore, using hybrid qubits of smaller values of $\alpha$ would improve the performance against photon loss as dephasing is mitigated. We shall show that adding these two features in building  $\ket{\mathcal{C}_\mathcal{L}}$ will lead to an improved $\eta_{\rm th}$ over HTQC.   
For brevity, we coin this new scheme as hybrid-qubit-based topological QC with postselection and $n$-BSM EO (PHTQC-$n$).

\begin{figure}[t]
\includegraphics[width=.9\columnwidth]{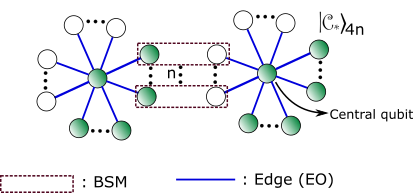}
\caption{A $4n$-arm star cluster state, $\ket{\mathcal{C}_\ast}_{4n}$ has a central qubit and $4n$ number of surrounding arm qubits. The arm qubits are entangled to the central qubit via edges. An edge between the two central qubits is created by performing multiple BSMs to which $n$ arm qubits of each $\ket{\mathcal{C}_\ast}_{4n}$ are inputs as shown. In the process, the BSMs are performed in a sequence until one of them succeeds or all the $n$-arm qubits are exhausted. As the BSMs in linear optics are probabilistic, using multi-BSM EOs make the edge creation near-deterministic. Thus, a $\ket{\mathcal{C}_\mathcal{L}}$ is built by entangling the $\ket{\mathcal{C}_\ast}_{4n}$'s to their four nearest neighbors.}
\label{fig:nHBSM}
\end{figure}

In PHTQC-$n$, implementing  $\ket{\mathcal{C}_\mathcal{L}}$ commences with the creation of a $4n$-arm star cluster state $\ket{\mathcal{C}_\ast}_{4n}$, where $n=1,2,3\dots$. The state represented by $\ket{\mathcal{C}_\ast}_{4n}$ has a central qubit and $4n$ number of surrounding arm qubits. The arm qubits are entangled to the central qubit, which are represented by the edges as shown in Fig.~\ref{fig:nHBSM}.  Here, unlike HTQC, we employ {\it postselection} in generation of $\ket{\mathcal{C}_\ast}_{4n}$ so that it has all the edges between the central qubit and the arm qubits intact. 
Further, the cluster state $\ket{\mathcal{C}_\mathcal{L}}$ is formed by entangling the central qubits of $\ket{\mathcal{C}_\ast}_{4n}$'s. This EO or creation of edges between the central qubits is achieved by performing multiple BSMs to which $n$-arm qubits of each $\ket{\mathcal{C}_\ast}_{4n}$ are inputs as shown in the Fig.~\ref{fig:nHBSM}. Thus, only the central qubit of $\ket{\mathcal{C}_\ast}_{4n}$ stays  in  $\ket{\mathcal{C}_\mathcal{L}}$. It is important to note that we perform up to $n$  HBSMs in a sequence until one succeeds or all are exhausted.

\begin{figure}[t]
\includegraphics[width=.7\columnwidth]{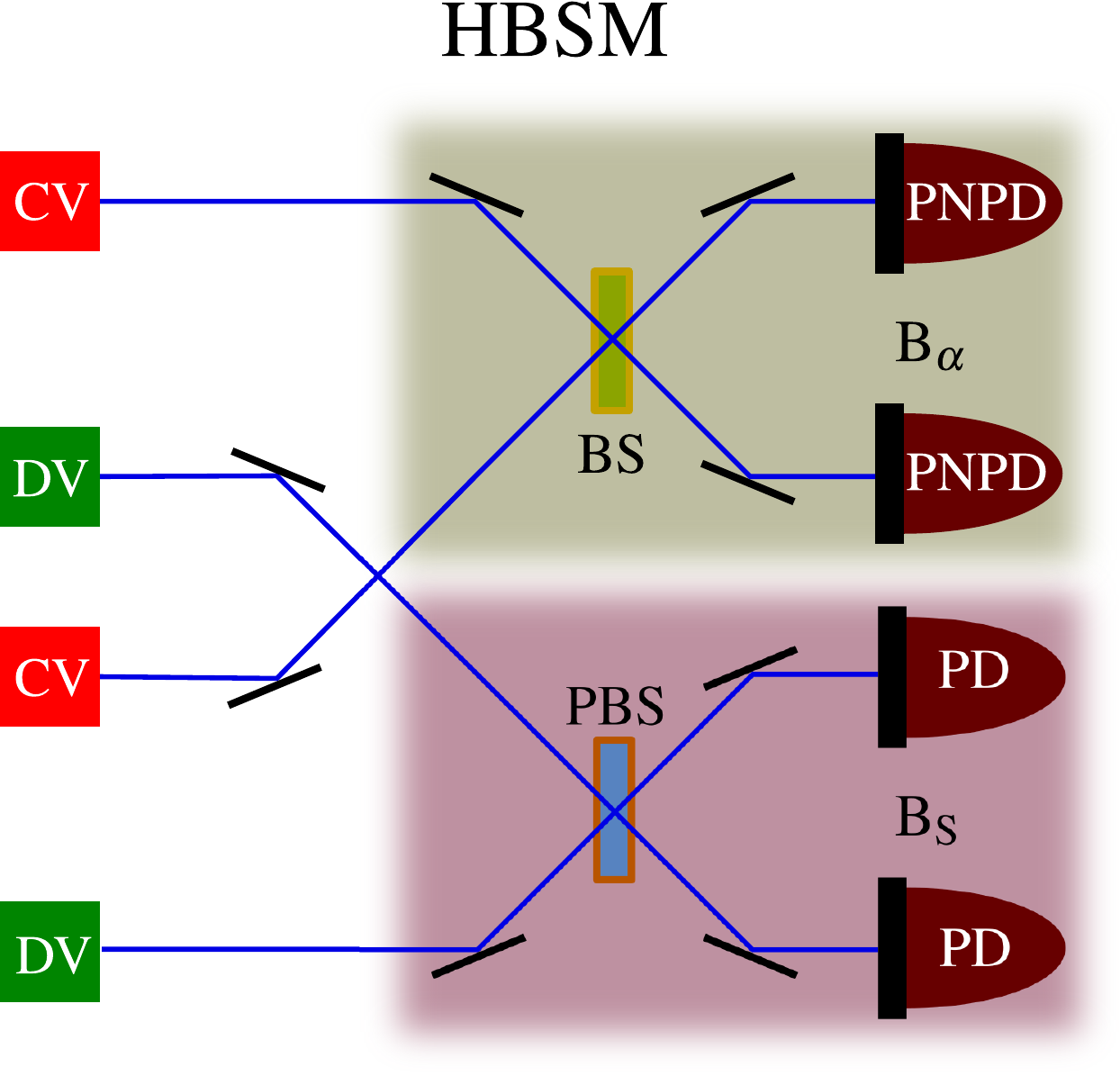}
\caption{$\Ba$ acts on the CV modes and fails when neither of the two PNPDs click. Its failure rate on the hybrid qubits is $\E{-2\alpha^2}$~\cite{LJ13}. 
${\rm B_S}$ acts on the DV modes and is successful with probability $1/2$ only when both the PDs click.  A HBSM fails only when both the $\Ba$ and ${\rm B_S}$ fail. Thus the failure rate of a HBSM is  $\E{-2\alpha^2}/2$.
}
\label{fig:bsm}
\end{figure}

BSM on hybrid-qubits is a composite of two BSM operations:  ${\rm B_S}$ and $\Ba$ acting on DV and CV parts of hybrid-qubit, respectively as shown in Fig.~\ref{fig:bsm}. We shall henceforth refer to such a composite as a {\it hybrid} BSM (HBSM), and its failure rate drastically approaches to zero with an increasing value of $\alpha$~\cite{LJ13,OTJ19}. 
 The measurement $\Ba$ comprises a beam splitter (BS) and two photon-number parity detectors (PNPD), whereas ${\rm B_s}$ has a polarizing beam splitter (PBS), two photo-detectors (PD). 
For more details refer to Ref.~\cite{OTJ19}.
A HBSM fails when both the constituent modules $\Ba$ and ${\rm B_s}$ fail. More precisely, the  failure rate of $\Ba$ at which no click is registered on the PNPDs  is $\E{-2\alpha^2}$ and that of ${\rm B_s}$ at which only one detector or none clicks  is $1/2$~\cite{OTJ19}. Thus, the failure rate of HBSM turns out to be $\E{-2\alpha^2}\!\!/2$. It is important to note that a HBSM failure is heralded so that the knowledge is available for decoding during QEC, postselection and multi-HBSM EOs.

As  BSMs  are not deterministic in linear optics, failures of EOs leave the corresponding edges between the qubits of $\ket{\mathcal{C}_\mathcal{L}}$ missing. This problem of missing edges can be addressed by transforming them to missing qubits of $\ket{\mathcal{C}_\mathcal{L}}$~\cite{LDSB10}. Then, topological QEC is carried out as detailed in  Sec.~\ref{sec:mbtqc}. When the missing fraction of the lattice qubits is  0.249 or more,  $\ket{\mathcal{C}_\mathcal{L}}$ can not support FTQC~\cite{LZ98}. In REf.~\cite{OTJ19}, HTQC overcomes this problem by using hybrid qubits on which BSM is near-deterministic due to larger value of $\alpha$. Thus, having only four arms in the star cluster  ($\ket{\mathcal{C}_\ast}_4$) suffices. But, smaller values of $\alpha$ would be appreciable for a better $\eta_{\rm th}$ (see Sec.~\ref{sec:noise}). Alternatively, when the BSM is probabilistic, Refs.~\cite{ LDSB10, FT10, HFJR10, LHMB15} tackle the problem by having multi-BSM EOs to improve success rate of edge creation. 
Similarly, in PHTQC-$n$ we employ $n$-HBSM EOs so that we can afford smaller values of $\alpha$ and still have edge creation near-deterministically.  This requires a larger $n$ value.

In this multi-BSM EO approach, BSMs are performed  sequentially  until one succeeds or all the $n$ arm qubits exhaust as shown in Fig.~\ref{fig:nHBSM}. With this strategy the incurred resources, in terms of both qubits and BSM trials, grow exponentially as the success rate of BSM falls. Moreover, once a BSM is successful, all other arm qubits must be removed using $Z$ measurements~\cite{FT10} as even number of successful BSMs correspond to removal of the edge.  Additionally, one must employ {\it active switching} for sequencing the multiple BSMs.  Switching is also known to be a major contributor for photon loss~\cite{LHMB15}. However, if the success rate of the BSM is high, the complexity of the switching circuit, and hence the photon loss can be reduced. 
 In this work we study in detail how PHTQC-$n$  performs against photon loss inspite of the apperant former disadvantage.

\subsection{Measurements on hybrid qubits of  $\ket{\mathcal{C}_\mathcal{L}}$}
The measurements on the hybrid qubits of $\ket{\mathcal{C}_\mathcal{L}}$ for topological FTQC can be achieved in two ways; either by measuring the DV or CV modes. Measurements on the DV mode are accomplished by detecting the polarization of the photons in respective basis. For CV modes $X$-measurement can be achieved by detections on  PNPDs, and  $Z$-measurement by homodyne detection in the displacement-quadrature~\cite{MR11}.
Measurements in the $(X\pm Y)/\sqrt{2}$ basis can be achieved by using the displacement operation in photon counting \cite{ITW+17} of the CV modes.  However, measurements on the DV modes alone are suffient for carrying out PHTQC-$n$.

\subsection{In-line and off-line processes}
The process of building  $\ket{\mathcal{C}_\mathcal{L}}$ consists of two stages: offline and inline stages. During the offline stage two types of 3-hybrid-qubit cluster states are generated probabilistically using $\ket{\Psi_\alpha}$ in Eq.~\eqref{eq:raw} as raw resources~\cite{OTJ19}. As the offline process is probabilistic, postselection is a necessity.  Once there is a continuous supply of the offline resource states (3-hybrid-qubit cluster states), the inline stage commences  by creating copies of $\ket{\mathcal{C}_\ast}_{4n}$ and entangling them to form lattice qubits and edges. Using $4n-2$ HBSMs on $4n-1$ offline resource states,  $\ket{\mathcal{C}_\ast}_{4n}$ can be generated by post-selecting over all the successful HBSMs. For example, $\ket{\mathcal{C}_\ast}_4$ can be  generated using two HBSMs on the off-line resource states as shown in Fig.~\ref{fig:star}. Further, $\ket{\mathcal{C}_\ast}_8$ can be generated using two $\ket{\mathcal{C}_\ast}_4$ and a 3-hybrid qubit cluster state, and two HBSMs as shown in Fig.~\ref{fig:8arm}~; a total of 6 HBSMs are required. 


\section{Generation of Star cluster state}
\label{sec:star}
First, we describe in detail how to create $|\mathcal{C}_\ast\rangle_4$ using offline resource states and HBSMs. This procedure is similar to that in HTQC, but involves postselection. Subsequently, we shall show how to extend the procedure to create  $|\mathcal{C}_\ast\rangle_{8}$, and more generally  to $|\mathcal{C}_\ast\rangle_{4n}$ with $n>2$. 

\begin{figure}[t]
\includegraphics[width=1\columnwidth]{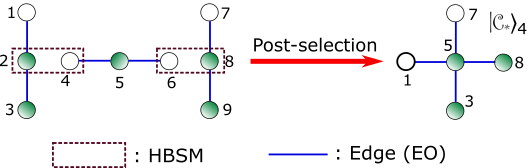}
\caption{The 3-hybrid-qubit offline resource state with an unfilled circle represents $\ket{\mathcal{C}_3}$ while that with two the $\ket{\mathcal{C}_{3^\prime}}$. Success of both HBSMs create a 4-arm star-cluster state $\ket{\mathcal{C}_\ast}$ and other cases leads to  undesired states as shown in Fig.~1(b) of Ref.~\cite{OTJ19}. In this work, we post-select on both HBSMs being successful and other cases are discarded. 
}
\label{fig:star}
\end{figure}

A $|\mathcal{C}_\ast\rangle_4$ is created using two kinds of offline resource states and two HBSMs as shown in  Fig.~\ref{fig:star}. The two offline resource states have the form 
\bea
\label{eq:3clus}
\ket{\mathcal{C}_3}&=&
\frac{1}{2}\big(\ket{\alpha,\alpha,\alpha}  \ket{\textsc{h},\textsc{h},\textsc{h}}+ \ket{\alpha,\alpha,-\alpha}\ket{\textsc{h},\textsc{h},\textsc{v}} \nonumber\\
&&+\ket{-\alpha,-\alpha,\alpha}  \ket{\textsc{v},\textsc{v},\textsc{h}} - \ket{-\alpha,-\alpha,-\alpha}\ket{\textsc{v},\textsc{v},\textsc{v}}\big)\,,\nonumber\\
\ket{\mathcal{C}_{3^\prime}}&=&
\frac{1}{\sqrt{2}}\big(\ket{\alpha,\alpha,\alpha}  \ket{\textsc{h},\textsc{h},\textsc{h}}+ \ket{-\alpha,-\alpha,-\alpha}\ket{\textsc{v},\textsc{v},\textsc{v}}\big).\nonumber\\
\eea
One can verify that  $\ket{\mathcal{C}_3}$ is the result of a Hadamard on the first qubit of the 3-qubit linear cluster state  ${\rm CZ}_{2,1}{\rm CZ}_{2,3}\ket{+}_1\ket{+}_2\ket{+}_3$. On the other hand, $\ket{\mathcal{C}_{3^\prime}}$ is due to a Hadamard on the first and third qubits of this 3-qubit linear cluster state.

It is important to note that the hybrid-qubit-based scheme faciliates the generation of 3-qubit cluster states using only linear optics with practical values of $\alpha$. Although coherent superposition states, $\ket{\alpha}\pm\ket{-\alpha}$ (up to normalization) also support near-deterministic BSM, by using only linear optics it is not possible to generate high-fidelity  3-qubit cluster states like $\frac{1}{\sqrt{2}}\big(\ket{\alpha,\alpha,\alpha}+ \ket{\alpha,\alpha,-\alpha}+\ket{-\alpha,-\alpha,\alpha} - \ket{-\alpha,-\alpha,-\alpha}\big)$ with practical values of $\alpha$. For example, the scheme in Ref.~\cite{MR11} needs $\alpha\approx10$ for a fidelity of $\sim0.9$, which is very low for QEC on  $\ket{\mathcal{C}_\mathcal{L}}$. For building a $\ket{\mathcal{C}_\mathcal{L}}$ suitable for topological FTQC, one needs initial coherent superposition states of very large $\alpha$~~\cite{MR11}.  Moreover, when  $\alpha$ is large  dephasing in the presence of photon loss is very strong on the qubits of $\ket{\mathcal{C}_\mathcal{L}}$, resulting in failure of QEC. 
As such, non-linear optical schemes like a cavity QED generation scheme~\cite{MSVR08} are necessary to build a suitable cluster state under those situations. 
On the other hand, we will demonstrate that using hybrid qubits of amplitude $\alpha<1$, it is possible to build a sufficiently good $\ket{\mathcal{C}_\mathcal{L}}$ for topological FTQC.

As shown in  Fig.~\ref{fig:star}, two $\ket{\mathcal{C}_3}$'s and a $\ket{\mathcal{C}_{3^\prime}}$  are initialized as  $\ket{\mathcal{C}_3}_{1,2,3}\otimes\ket{\mathcal{C}_{3^\prime}}_{4,5,6}\otimes\ket{\mathcal{C}_3}_{7,8,9}$,  and  two HBSMs act on the modes  2, 4 and 6, 8 respectively. We note that the HBSM components, $\Ba$ and ${\rm B_s}$  respectively act on CV and DV modes of the hybrid qubits.
When the HBSMs acting on modes 2, 4 and 6, 8 are successful, that is either $\Ba$ or ${\rm B_s}$ succeedes, and say $\Ba$ has an outcome corresponding to the Bell state $\ket{\psi^+}$, then the resulting state would be 
\bea
\label{eq:star}
\ket{\mathcal{C}_\ast}_4&=&\ket{\alpha,\alpha,\alpha,\alpha,\alpha}_{1,3,5,7,8}  \ket{\textsc{h},\textsc{h},\textsc{h},\textsc{h},\textsc{h}}_{1,3,5,7,8}\nonumber\\
&&+\ket{\alpha,\alpha,\alpha,\alpha,-\alpha}_{1,3,5,7,8}  \ket{\textsc{h},\textsc{h},\textsc{h},\textsc{h},\textsc{v}}_{1,3,5,7,8}\nonumber\\ 
&&+\ket{\alpha,-\alpha,\alpha,\alpha,\alpha}_{1,3,5,7,8}  \ket{\textsc{h},\textsc{v},\textsc{h},\textsc{h},\textsc{h}}_{1,3,5,7,8}\nonumber\\ 
&&+\ket{\alpha,-\alpha,\alpha,\alpha,-\alpha}_{1,3,5,7,8}  \ket{\textsc{h},\textsc{v},\textsc{h},\textsc{h},\textsc{v}}_{1,3,5,7,8}\nonumber\\ 
&&+\ket{-\alpha,\alpha,-\alpha,-\alpha,\alpha}_{1,3,5,7,8}  \ket{\textsc{v},\textsc{h},\textsc{v},\textsc{v},\textsc{h}}_{1,3,5,7,8}\nonumber\\ 
&&-\ket{-\alpha,\alpha,-\alpha,-\alpha,-\alpha}_{1,3,5,7,8}  \ket{\textsc{v},\textsc{h},\textsc{v},\textsc{v},\textsc{v}}_{1,3,5,7,8}\nonumber\\ 
&&-\ket{-\alpha,-\alpha,-\alpha,-\alpha,\alpha}_{1,3,5,7,8}  \ket{\textsc{v},\textsc{v},\textsc{v},\textsc{v},\textsc{h}}_{1,3,5,7,8}\nonumber\\ 
&&+\ket{-\alpha,-\alpha,-\alpha,-\alpha,-\alpha}_{1,3,5,7,8}  \ket{\textsc{v},\textsc{v},\textsc{v},\textsc{v},\textsc{v}}_{1,3,5,7,8}.\nonumber\\
\eea
Upon getting other three different possible  outcomes for $\Ba$'s and two for ${\rm B_I}$,  the resulting $\ket{\mathcal{C}_\ast}_4$ would be equivalent to the one in  Eq.~\eqref{eq:star} up to local Pauli rotations. This can be handled by updating the {\it Pauli frame} without the need for any feed-forward optical operations.

A desired $|\mathcal{C}_\ast\rangle_4$ with edges connecting the central qubit to all the arm qubits is generated only when both HBSMs are successful. In other cases, that is when one of the HBSMs fails or both, the resulting states are distorted with edges  misplaced between surrounding qubits (refer to Fig.~1(b) of Ref.~\cite{OTJ19}). To see this, suppose that the  HBSM acting on modes 2 and 4 of  the initialized state  $\ket{\mathcal{C}_3}_{1,2,3}\otimes\ket{\mathcal{C}_{3^\prime}}_{4,5,6}\otimes\ket{\mathcal{C}_3}_{7,8,9}$  fails and the other succeeds. The resulting state is 
\bea
&&\big(\ket{\alpha,\alpha}_{1,3}  \ket{\textsc{h},\textsc{h}}_{1,3}+\ket{\alpha,-\alpha}_{1,3}  \ket{\textsc{h},\textsc{v}}_{1,3}+\ket{-\alpha,\alpha}_{1,3}  \ket{\textsc{v},\textsc{h}}_{1,3}\nonumber\\
&&+\ket{-\alpha,-\alpha}_{1,3}  \ket{\textsc{v},\textsc{v}}_{1,3}\big)\otimes \ket{\mathcal{C}_3}_{5,7,8}.
\eea
We can observe that there are no edges (no entanglement) from the central qubit (mode 5) to qubits 1 and 3. Rather, there is a misplaced edge between qubits 1 snd 3. We refer to this misplaced edges as diagonal edges due to its geometric appearance (see Fig.~1(b) of Ref.~\cite{OTJ19}). This leads to distortion of the lattice geometry and stabilizer structure. Each failure of HBSM results in two missing edges and an undesired diagonal edge  in a $\ket{\mathcal{C}_\ast}_4$.  Contrary to Ref.~\cite{OTJ19} that utilizes distorted star cluster states, here we employ postselection to choose $\ket{\mathcal{C}_\ast}_4$ with intact edges. 
Therefore, the resulting  $\ket{\mathcal{C}_\mathcal{L}}$ would be free of diagonal edges. 
As avoiding the diagonal edges leads to lesser number of missing lattice-edges,  postselection results in a lattice with much fewer missing qubits and a better tolerance against dephasing.

\subsection{Star cluster state with more than 4-arm qubits}
 Increasing the number of arms provides an opportunity to repeat the HBSM operations when the previous ones fail.  
 The bottleneck here is that as the number of arms goes up, the success rate of HBSMs fall (as $\alpha$ correspondingly decreases) and there is a growing complexity in the switching circuit for postselection. 
It is also known that switching adds to photon loss which would be detrimental for $\eta_{\rm th}$. 
For those reasons, we restrict to utilizing only $\ket{\mathcal{C}_\ast}_{8}$ and  $\ket{\mathcal{C}_\ast}_{12}$.

\begin{figure}[t]
\includegraphics[width=.95\columnwidth]{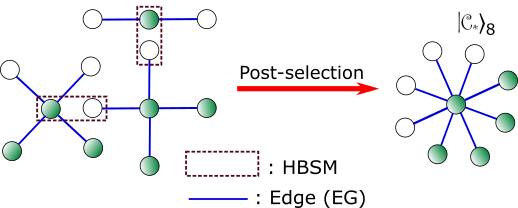}
\caption{(a) An 8-arm start cluster state $\ket{\mathcal{C}_\ast}$ can be created by entangling two 4-arm $\ket{\mathcal{C}_\ast}$'s and a three qubit-cluster state $\ket{\mathcal{C}_{3^\prime}}$ with two HBSMs. postselection is employed to obtain intact states. Further, copies of 8-arm $\ket{\mathcal{C}_\ast}$ are entangled to form a $\ket{\mathcal{C}_{\mathcal{L}}}$.}
\label{fig:8arm}
\end{figure}

A $\ket{\mathcal{C}_\ast}_8$  can be created by entangling two $\ket{\mathcal{C}_\ast}_4$'s and a $\ket{\mathcal{C}_{3^\prime}}$ with two HBSMs as described in  Fig.~\ref{fig:8arm}, where postselection is carried out over the successful HBSMs. Similar to  the case of $\ket{\mathcal{C}}_4$, one can explicitly work out and show that the process in Fig.~\ref{fig:8arm} would result in  $\ket{\mathcal{C}_\ast}_8$. By the same token, one can generate  $\ket{\mathcal{C}_\ast}_{12}$ by entangling  $\ket{\mathcal{C}_\ast}_{8}$,  $\ket{\mathcal{C}_\ast}_4$ and $\ket{\mathcal{C}_{3^\prime}}$ with two HBSMs.

After building $|\mathcal{C}_\mathcal{L}\rangle$ using $\ket{\mathcal{C}_\ast}_{4n}$'s  with postselection, both QEC and gate operations on the topological states of the lattice are executed  by measuring the hybrid qubits individually. The measurements, in principle, transfers the state on a layer to the next in a similar manner as in teleportation. 
In practice, two layers of $|\mathcal{C}_\mathcal{L}\rangle$ suffice at any instant. The third-dimension of $|\mathcal{C}_\mathcal{L}\rangle$ is a happening in time~\cite{RHG06}.

\section{Noise model}
\label{sec:noise}
The predominant errors in optical quantum computing models originate from  photon loss~\cite{RP10}. In this section, we study the effect of the photon loss on hybrid qubits and in turn on  PHTQC-$n$.   The action of  photon-loss channel $\mathcal{E}$ on a hybrid qubits initialized to the  state $\rho_0=\ket{\Psi_\alpha}\langle\Psi_\alpha|$ gives~\cite{LJ13}
\bea
\label{eq:rho}
\mathcal{E}(\rho_0)&=&\frac{(1-\eta)}{2}\bigg(|\alpha^\prime,\textsc{h}\rangle\langle\alpha^\prime,\textsc{h}|+|-\alpha^\prime,\textsc{v}\rangle\langle-\alpha^\prime,\textsc{v}|\nonumber\\
&+&e^{-2\eta\alpha^2}\big(|\alpha^\prime,\textsc{h}\rangle\langle-\alpha^\prime,\textsc{v}|+|-\alpha^\prime,\textsc{v}\rangle\langle\alpha^\prime,\textsc{h}|\big)\bigg)\nonumber\\
&+&\frac{\eta}{2}\bigg(\big(|\alpha^\prime\rangle\langle\alpha^\prime|+|-\alpha^\prime\rangle\langle-\alpha^\prime|\big)\otimes\ket{0}\langle0|\bigg),\nonumber\\
&=&(1-\eta)\bigg(\frac{1+\E{-2\eta\alpha^2}}{2}\ket{\Psi_{\alpha^\prime}^+}\langle\Psi_{\alpha^\prime}^+|\nonumber\\
&&+\dfrac{1-\E{-2\eta\alpha^2}}{2}\ket{\Psi_{\alpha^\prime}^-}\langle\Psi_{\alpha^\prime}^-|\bigg)\nonumber\\
&&+\frac{\eta}{2}\big(\ket{\Phi_{\alpha^\prime}^+}\langle\Phi_{\alpha^\prime}^+|+\ket{\Phi_{\alpha^\prime}^-}\langle\Phi_{\alpha^\prime}^-|\big),
\eea
 where $\ket{\Psi_{\alpha^\prime}^\pm}=(\ket{\alpha^\prime,\textsc{h}}\pm\ket{-\alpha^\prime,\textsc{v}})/\sqrt{2},~ \ket{\Phi_{\alpha^\prime}^\pm}=\ket{0}\otimes(\ket{\alpha^\prime}\pm\ket{-\alpha^\prime})/\sqrt{2}$, and $\alpha^\prime=\sqrt{1-\eta}\alpha$ with $\eta$ being the photon-loss rate that arises from imperfect sources and detectors, absorptive optical components and storage. 
 The effect of photon loss on hybrid-qubits is to introduce phase-flip errors $Z$ and  diminishes the amplitude $\alpha$ to  $\alpha^\prime$, which consequently lowers the success rate of HBSMs. Its also forces the hybrid qubit state to leak out of the computational basis,  $\{\ket{0}_{\rm L},\ket{1}_{\rm L}\}$.

 From Eq.~\eqref{eq:rho} one can deduce that the dephasing rate is 
\be
\label{eq:pz}
p_z=(1-\eta)\frac{1-\E{-2\eta\alpha^2}}{2}+\frac{\eta}{2}=\frac{1}{2}[1-(1-\eta)\,\E{-2\eta\alpha^2}],
\ee
 which increases with the value of $\alpha$ for a given $\eta$. Thus, for a fixed value of $\eta$, we face a trade-off between the desirable success rate of HBSM and the detrimental effects of dephasing with increasing value of $\alpha$.  
Owing to photon loss,  the failure  rate of a HBSM reads
\be
p_f=\frac{1}{2}(1-\eta)\E{-2\alpha^{\prime 2}}+\eta\E{-2\alpha^{\prime 2}}= \frac{1}{2}(1+\eta)\E{-2\alpha^{\prime 2}}.
\ee
In the above equation,  the first term originates from the attenuation of CV part, while the second from both CV attenuation and DV loss. We point out that like DV optical schemes \cite{HFJR10}, photon loss does not necessarily imply lattice-qubit loss in PHTQC-$n$. The probability $\langle0,0|\mathcal{E}(\rho_0)|0,0\rangle=\eta \E{-\alpha^{\prime 2}}$ that photon loss leading to lattice-qubit loss for $\eta\sim10^{-3}$ is much smaller than the HBSM failure rate, $p_f$ and can be neglected.

\section{Simulation of QEC }
\label{sec:sim}
To simulate  QEC on the $|\mathcal{C}_\mathcal{L}\rangle$ of hybrid-qubits with missing edges and dephasing noise, we use the software package  AUTOTUNE~\cite{FWMR12}. It 
offers a wide range of options for noise models and their customization to suit our scheme. Most importantly, it allows for the simulation  of QCE when the qubits are missing. We obtain  results by exploiting this feature {\it via} mapping  missing edges to  missing qubits~\cite{AAG+18}. 

AUTOTUNE  uses the circuit model (where qubits are initialized in the $\ket{+}$ state and CZ operations are applied to create entanglement between the appropriate qubits) to simulate the error propagation during the formation of
  $|\mathcal{C}_\mathcal{L}\rangle$.  Here, we detail how noise in PHTQC-$n$ (which 
employs techniques different from the circuit model for building $|\mathcal{C}_\mathcal{L}\rangle$) can be simulated using AUTOTUNE. 
As explained in  Sec.~\ref{sec:phtqc}, only the central hybrid qubit of $\ket{\mathcal{C}_\ast}_{4n}$ remains in the lattice and the arm qubits are utilized by the HBSMs. All the   hybrid qubits of $\ket{\mathcal{C}_\ast}_{4n}$ suffer from dephasing  of rate $p_Z$ in Eq.~\eqref{eq:pz} due to photon loss. The action of  HBSMs transfer noise on the arm qubits to the central qubits~\cite{OTJ19}. Thus, the central qubits accumulate additional noise due to the HBSMs.  The role of a HBSM in creating  edges between the central qubits of $|\mathcal{C}_\ast\rangle_{4n}$, is equivalent to that of a CZ in the circuit model for building $|\mathcal{C}_\mathcal{L}\rangle$. So, the action of HBSMs under noise in PHTQC can be simulated by noisy CZs in  AUTOTUNE.  Once a noisy $|\mathcal{C}_\mathcal{L}\rangle$ is simulated, the QEC proceeds the same way for both pictures.  AUTOTUNE also allows for the simulation of  noise introduced during the initialization of qubits that mimics a noisy $|\mathcal{C}_\ast\rangle_{4n}$ and subsequent error propagation through the action of HBSMs. Other operations in AUTOTUNE  that are not relevant to us are set to be noiseless.

More specifically, noise from a HBSMs is simulated as  noise introduced by a CZ described by the Kraus operators  $\{ \sqrt{(1-2p_Z)}~ I\otimes I,~~\sqrt{p_Z}~  Z\otimes I,~~ \sqrt{p_Z} ~ I\otimes Z \}$~\cite{OTJ19}. In PHTQC-$n$, HBSMs act up to $n$ times to create an edge between two central qubits. 
The rate of dephasing added by $n$ HBSMs on the central qubits is $1-(1-p_Z)^n$. In the limit $p_Z<<1$, this amounts to $np_Z$.
 Accordingly, the noise corresponding to $n$ HBSMs would have the following Kraus operators:  $\{ \sqrt{(1-2np_Z)}~ I\otimes I,~~\sqrt{np_Z}~  Z\otimes I,~~ \sqrt{np_Z} ~ I\otimes Z \}$. 
It is easy to show that the average number of HBSMs needed to create an edge is 
\be
n_\mathrm{avg}=\sum^\infty_{m=0}(1-p_f)p_f^m(m+1)=\frac{1}{1-p_f},
\ee
 where $p_f\ll1-\frac{1}{n}$ if an edge between two central hybrid qubits should typically be created with $n$ HBSMs.
 Thus, in PHTQC-$n$,   a noisy entangling operation is described by the set of Kraus operators: $\{ \sqrt{(1-2n_{\rm avg}p_Z)}~ I\otimes I,~~\sqrt{n_{\rm avg}p_Z}~  Z\otimes I,~~ \sqrt{n_{\rm avg}p_Z} ~ I\otimes Z \}$. This description also accounts for dephasing in the active switiching process.
AUTOTUNE can also allows to simulate instances when no gate actions happen, but qubits suffer loss and dephasing.  We assume that the postselection takes place in this instance. This shall allow us to account for photon loss and dephasing  during the switching process in the simulation.

Further, the QEC simulation on a faulty $\ket{\mathcal{C}_\mathcal{L}}$ begins by making $X$-basis measurements on the noisy hybrid qubits. We again introduce dephasing of  rate $p_Z$ on the hybrid qubits waiting to undergo measurement. The $X$-measurement outcomes used for syndrome extraction during QEC could be  erroneous. This error rate is also set at $p_Z$. Due to photon loss, the hybrid qubits may leak out of the logical basis which makes measurements on such DV modes impossible. This  leakage error too is assigned the same  rate of $p_Z$. As $p_Z>\eta$, the assignment will only overestimates the leakage error of rate $\eta$.



\section{Results on photon loss threshold}
\label{sec:result}
\begin{figure}[h!]
\includegraphics[width=1\columnwidth]{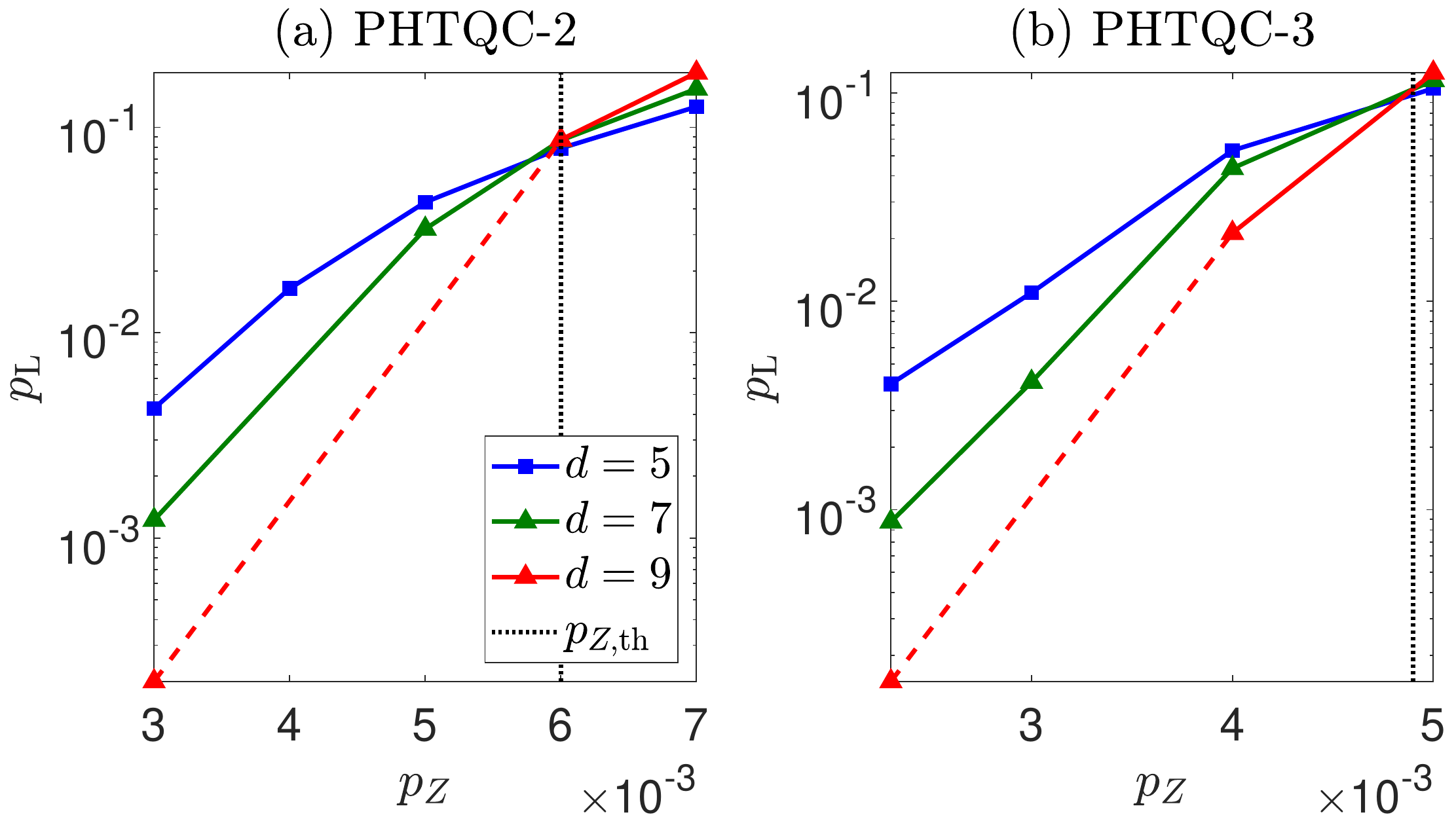}
\caption{\label{fig:qm/complexfunctions} Logical error rate $p_{\rm L}$ is plotted against the dephasing rate $p_Z$ for PHTQC-2 and PHTQC-3 
 of code distances $d=5,7,9$. The intersecting point of these curves corresponds to the threshold dephasing rate $p_{Z,\rm th}$. The plots correspond to the qubit loss rate $p_{\rm loss}=0.03$.}
\label{fig:result}
\end{figure}

The {\it logical error}  rate $p_{\rm L}$ is determined against  value of  $p_Z$ for $\ket{\mathcal{C}_\mathcal{L}}$ of  code distances $d$ using  AUTOTUNE.  This calculation is repeated for various values of lattice-qubit loss rate $p_{\rm loss}$, which correspond to different values of $\alpha$.  The intersection point of the curves corresponding to various $d$'s is the threshold dephasing rate $p_{Z,\rm th}$ as marked in Fig.~\ref{fig:result}.   The photon loss threshold $\eta_{\rm th}$ is determined using Eq.~\eqref{eq:pz} by replacing $p_Z$ with $p_{Z,\rm th}$.

From Fig.~3(b) of Ref.~\cite{OTJ19}, we estimate that PHTQC-$n$ would also perform best around $p_{\rm loss}=0.03$ as both schemes use the same error model. Thus we simulate the QEC for both PHTQC-2 and PHTQC-3 with $p_{\rm loss}=0.03$.  To determine the values of $\alpha$ required for PHTQC-$n$, we map $p_{\rm loss}$ to $p_f$ as detailed in the following.  

Each qubit in $\ket{\mathcal{C}}_\mathcal{L}$ is associated with four edges created by $n$-HBSM. When all HBSMs fail the edge between the lattice qubits will be missing.  In this scenario, either of the qubits is removed with equal probability thereby mapping a missing edge to a missing qubit~\cite{AAG+18}. In PHTQC-$n$, probability of having a missing edge is $p_f^n$. A qubit is removed when more than one of the associated edges is missing giving us
\be
p_{\rm loss}=1-\big(1-\frac{1}{2}p_f^n\big)^4.
\label{eq:ploss}
\ee

After inserting the values of $p_{\rm loss}$ and $n$ into Eq.~\eqref{eq:ploss}, we find that PHTQC-2 and PHTQC-3 require $\alpha$ of values 0.84 and 0.6, respectively. From the simulation result for PHTQC-2, as shown in Fig.~\ref{fig:result}(a), we have  $p_{Z,{\rm th}}=0.006$, which translates to $\eta_{\rm th}=5\times10^{-3}$ with the aid of Eq.~\eqref{eq:pz}. Similarly, from  Fig.~\ref{fig:result}(b) we have $p_{Z,{\rm th}}=0.0049$ that results in $\eta_{\rm th}=5.7\times10^{-3}$ for PHTQC-3. 
These results imply that PHTQC-2 and PHTQC-3 schemes provide improved values of $\eta_{\rm th}$ over HTQC. Compared to other known optical QC schemes~\cite{DHN06,Cho07,HHGR10,LPRJ15,HFJR10,LRH08,LJ13}, we find that PHTQC-3 provides the highest $\eta_{\rm th}$   (when computational error rate is nonzero).

\section{Results on Resource overhead} 
\label{sec:resource}
\begin{figure}[t]
\includegraphics[width=.8\columnwidth]{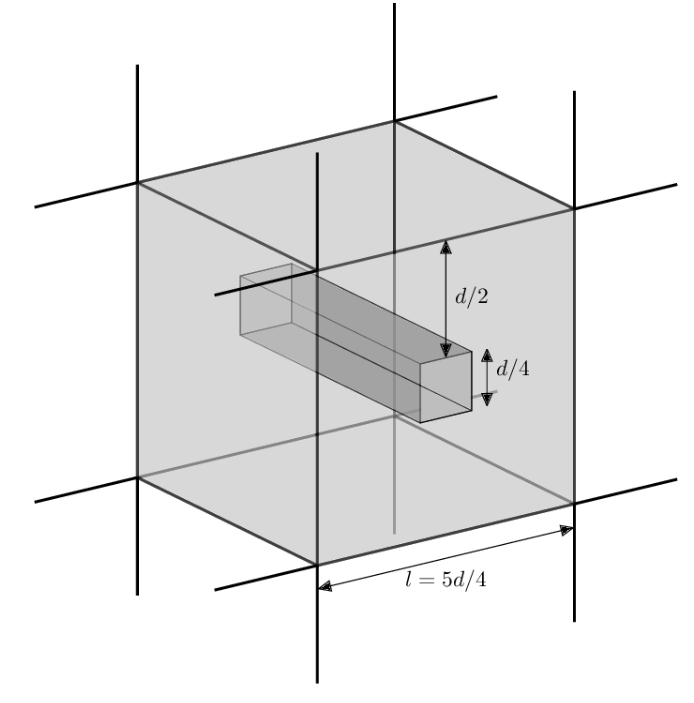}
\caption{ To achieve fault tolerance for gate operations, both the circumference of a defect and distance between two defects should be $d$. So, the defects are placed in the cubic lattice of side $l=5d/4$ as shown such that distance between the defects is $d$. Thus on average, a cubic lattice of volume $(5d/4)^3$ is required per fault tolerant gate operation. 
}
\label{fig:volume}
\end{figure}

In order to estimate the resource overhead per fault-tolerant topological gate operation, we count the average number  of hybrid qubits $N$ required to build  a cubic-fraction of the $\ket{\mathcal{C}}_\mathcal{L}$ of sufficiently large side $l$ determined by the target $p_{\rm L}$. 
As depicted in Fig.~\ref{fig:volume}, $l$ is determined such that the cubic-fraction can accommodate a defect of circumference $d$ so that there are no error chains encircling it. Also, the defect is separated by a distance $d$~\cite{WF14} from those in neighboring cubic-fractions to avoid a chain of errors connecting them. For this, the side of the cubic-fraction must be at least $l=5d/4$. By extrapolating the suppression of $p_{\rm L}$ with $d$, we determine the value of $d$ required to achieve the target $p_{\rm L}\approx10^{-15}$ using the following expression~\cite{WF14} (see also Fig.~\ref{fig:result})
\be
p_{\rm L}=\frac{b}{(\frac{a}{b})^\frac{d-d_b}{2}},
\label{eq:pl}
\ee
where $a$ and $b$ are the values of $p_{\rm L}$ corresponding to the second highest and the highest $d$ values chosen for our simulations. We determine $a$ and $b$ below half the threshold value, that is $p_{Z,{\rm th}}/2$.  
 Once  $d$ is determined, $N$ can be estimated as detailed in the subsequent subsection. 
We emphasise that  $N$ also depends upon the value of the $\eta$ at which  FTQC schemes operate, that is 
closer to  threshold they operate, more resources are consumed.

\subsection{Resource overhead for PHTQC-$n$}
Let us recall that two $\ket{\mathcal{C}_3}$'s,  a $\ket{\mathcal{C}_{3^\prime}}$ and two HBSMs are needed to create $\ket{\mathcal{C}_\ast}_4$.  The success rate of both HBSMs  is $(1-\frac{1}{2}e^{-2\alpha^{\prime2}})^2$. On average, $8/[(1-\E{-2\alpha^{\prime2}})^2]$ hybrid qubits are needed to create a $\ket{\mathcal{C}_3}$ or $\ket{\mathcal{C}_{3^\prime}}$.   
 Taking postselection into account, the average number of hybrid qubits in building  $\ket{\mathcal{C}_\ast}_4$ would be 
$24/[(1-\E{-2\alpha^{\prime2}})^2(1-\frac{1}{2}e^{-2\alpha^{\prime2}})^2]$. In general, a $\ket{\mathcal{C}_\ast}_{4n}$ 
can be created by entangling a $\ket{\mathcal{C}_\ast}_{4n-4}$, a $\ket{\mathcal{C}_\ast}_4$ and a $\ket{\mathcal{C}_{3^\prime}}$ using two HBSMs. A  $\ket{\mathcal{C}_\ast}_{4n-4}$
in turn requires $4n-6$ HBSMs, while $\ket{\mathcal{C}_\ast}_4$ needs two HBSMs. Therefore, a total of $4n-2$ HBSMs are used in creating  $\ket{\mathcal{C}_\ast}_{4n}$, which is formed from  $n$ number of $\ket{\mathcal{C}_\ast}_{4}$'s and $n-1$ $\ket{\mathcal{C}_{3^\prime}}$. 
On average, one needs 
\be
\bigg[\frac{24n}{(1-\E{-2\alpha^{\prime2}})^2}+\frac{8(n-1)}{(1-\E{-2\alpha^{\prime2}})^2}\bigg]\frac{1}{(1-\frac{1}{2}\E{-2\alpha^{\prime2}})^{4n-2}}\nonumber
\ee
hybrid qubits to synthesize $\ket{\mathcal{C}_\ast}_{4n}$.

\begin{table*}
\begin{tabular}{ |c| c| c| c| c| c|c| }\hline
 Scheme & QEC Code  &$\eta_{\rm th}$& $\eta$, computational error rate  & Resource  &$N$ for $p_{\rm L}= 10^{-6}$ & $N$ for $p_{\rm L}=10^{-15}$ \\ \hline
 OCQC & 7-qubit Steane code&$\it{4\times10^{-3}}$ & $4\times10^{-4},~4\times10^{-5}$ & Bell pair &$2.6\times10^{19}$& $7.1\times10^{24}$\\  
  PLOQC & 7-qubit Steane code& $\it{2\times10^{-3}}$  &  $4\times10^{-4},~4\times10^{-5}$ &Bell pair&$6.8\times10^{14}$  &$3.5\times10^{19}$\\
EDQC & Error detecting codes& $\it{1.57\times10^{-3}}$ &$1\times10^{-4},~1\times10^{-5}$ &Bell pair &$\mathcal{O}(10^{13})$&$\mathcal{O}(10^{16})$\\
CSQC & 7-qubit Steane code& $2.3\times10^{-4}$ & $8\times10^{-5},~1.97\times10^{-5}$&  CSS qubits&$2.1\times10^{11}$&$6.9\times10^{15}$\\
MQQC & 7-qubit Steane code &$1.7\times10^{-3}$  & $\mathcal{O}(10^{-4}),~\mathcal{O}(10^{-4})$ &Bell pair&$2.7\times10^{14}$&$1.4\times10^{19}$\\
HQQC & 7-qubit Steane code& $4.6\times10^{-4}$  & $\mathcal{O}(10^{-4}),~\mathcal{O}(10^{-4})$ &Hybrid qubits&$8.2\times10^{9}$&$2.3\times10^{12}$\\
TPQC &Topological&$\it{5.3\times10^{-4}}$  & 0,~$1\times10^{-3}~(5.3\times10^{-4},~0)$& Entangled photons& $>2\times10^9$ & $>4.2\times10^{10}$\\
HTQC &Topological &$3.3\times10^{-3}$  & $1.5\times10^{-3},~3\times10^{-3}$ &Hybrid qubits&$8.5\times10^5$ & $1.7\times10^7$\\
PHTQC-2 & Topological & $5\times10^{-3}$ & $2.4\times10^{-3},~3\times10^{-3}$  &Hybrid qubits& $ 1.1\times10^6$ & $1.8\times10^7$\\
{\bf PHTQC-3} & Topological & $\bf 5.7\times10^{-3}$ & $2.6\times10^{-3},~2.3\times10^{-3}$  &Hybrid qubits& $2.9\times10^7$&  $4.9\times10^8$\\ \hline
\end{tabular}
\caption{The table lists various  fault-tolerant optical QC schemes, the associated QEC codes, type of optical resource used, the optimal photon loss threshold $\eta_{\rm th}$ they provide and incurred resource overhead $N$. 
The resource overhead $N$ to attain the logical error rate $p_{\rm L}\sim10^{-6}(10^{-15})$ is calculated for operational values of the photon loss rate $\eta$ and computational error rate. 
It should be noted that $\eta_{\rm th}$'s claimed by OCQC,  PLOQC, EDQC and TPQC (in {\it italic}) are valid only for zero computational errors, which is unrealistic since photon losses would cause computational errors.
It is clear that  PHTQC-3 offer highest $\eta$ and computational error rate by an order of magnitude compared to other known optical QC schemes along with the best resource efficiency. Note that in PHTQC-$n$ with $n>{\bf 3}$ the value of $p_f$ is comparable to that in DV scheme~\cite{LHMB15}  and the usage of hybrid qubits offers no advantage in significantly reducing $N$. We hence only provide results  up to $n=3$.
}
\label{tab:para}
\end{table*}

As mentioned in Sec.~\ref{sec:phtqc}, each $\ket{\mathcal{C}_\ast}_{4n}$ appears as a single qubit in the final lattice $\ket{\mathcal{C}_\mathcal{L}}$.  This means the number of  $\ket{\mathcal{C}_\ast}_{4n}$'s needed to build a lattice of side $l$ is $6l^3$. 
Finally, the average number of hybrid qubits needed for building  $\ket{\mathcal{C}_\mathcal{L}}$ of side $l=5d/4$ in PHTQC-$n$  is
\be
N_n=\bigg[\frac{32n-8}{(1-\E{-2\alpha^{\prime2}})^2}\bigg]
\frac{125d^3}{64\left(1-\frac{1}{2}\E{-2\alpha^{\prime2}}\right)^{4n-2}},
\label{eq:n1}
\ee
which is more than that for HTQC.

  For PHTQC-2 with $\alpha=0.84$, from Fig.~\ref{fig:result}(a) we have $a\approx1.2\times10^{-3}$, 
$b\approx 2\times 10^{-4}$ and $d_{b}=9$ and using these values in Eq.~\eqref{eq:pl}, we find that $d \approx 15~(39)$ is needed to achieve $p_{\rm L}\sim10^{-6}~(10^{-15})$. This incurs  $N_2\approx1.1\times10^{6}~(1.8\times10^{7})$ hybrid qubits. 
For PHTQC-3 with $\alpha=0.6$, from Fig.~\ref{fig:result}(b) we have $a\approx8.5\times10^{-3}$, 
$b\approx 1.7\times 10^{-4}$ and $d_{b}=9$. As in the previous case, using these values  we find that $d \approx 16~(41)$ is needed to achieve $p_{\rm L}\sim10^{-6}~(10^{-15})$. This incurs  $N_3\approx2.9\times10^{7}~(4.9\times10^{8})$ hybrid qubits.

\section{Comparison}
\label{sec:compare}
We briefly present  the known linear optical FTQC schemes based on DV, CV and hybrid platforms and compare their performance parameters (tabulated in Tab.~\ref{tab:para}) with those of PHTQC-$n$.

 Reference \cite{DHN06} is one of the earliest works that determines the threshold region of $\eta$ and computational error rate, and performs resource estimation for linear optical QC.  The scheme uses optical {\it cluster states}~\cite{RB01} for FTQC (which we abbreviate as OCQC), built using entangled polarization photon pairs. 
 This scheme uses CSS QEC codes~\cite{NC10} coupled with telecorretion, where teleportation is used for  error-syndrome extraction for fault-tolerance. OCQC uses concatenation of QEC codes to attain  low values of $p_{\rm L}$. For example, 6(4) levels of error correction were employed to attain $p_{\rm L}\sim10^{-15}(10^{-6})$.  Unfortunately, resource overhead $N$ demanded by  OCQC (see Tab.~\ref{tab:para}) is too high for practical purpose and subsequent studies aimed to reduce it.

Later, Ref.~\cite{Cho07} used {\it error-detecting quantum state transfer} (EDQC) for optical FTQC. The underlying codes were capable of detecting errors in a way similar to  the scheme in Ref.~\cite{Kni05}, where QEC is shown to be  possible by concatenating different error detecting codes.   EDQC offers a smaller $\eta_{\rm th}$, but the value of $N$ could be reduced by many orders of magnitude compared to OCQC (refer to Tab.~\ref{tab:para}). 
Another scheme, namely the parity state linear optical QC (PLOQC) scheme~\cite{ HHGR10} that encodes multiple photons into a logical qubit in {\it parity state},
 provides a smaller $\eta_{\rm th}$, but an improved resource efficiency compared to OCQC. This scheme, similar to OCQC, uses CSS QEC codes and telecorrection. 
There also exists the multi-photon qubit QC (MQQC) scheme~\cite{LPRJ15} that
uses telecorrection based on CSS QEC code. See Tab.~\ref{tab:para} for the parameters of performance of MQQC.
Similar to OCQC,  schemes EDQC, PLOQC and MQQC need few levels of concatenation of QEC codes to attain target  $p_{\rm L}$ (refer to supplemental material of Ref.~\cite{OTJ19} for details). 
 Using DV optical platform a topological photonic QC (TPQC) scheme was proposed~\cite{HFJR10}. In Ref.~\cite{HFJR10},  photonic topological QC (TPQC) scheme operating on a DV optical platform was proposed.
Here, FTQC is performed on $\ket{\mathcal{C}_\mathcal{L}}$ built from a stream of entangled polarization photons. The value of $N$ for TPQC is calculated either for $\eta=0$ or zero computational error rate (only those cases are considered in the Ref.~\cite{HFJR10}). When both the parameters are nonzero, $N$ would in principle be much larger.

The coherent-state quantum computation (CSQC)~\cite{JK02,Ralph03,LRH08} uses the following set of coherent states $\{\ket{\alpha},\ket{-\alpha}\}$ as the logical basis for CV qubits.  CSQC also executes telecorrection  for tolerance against photon-loss and computational errors~\cite{LRH08}. 
In this CV scheme, superpositions of superposition states, $\ket{\alpha}\pm \ket{-\alpha}$ (up to normalization) \cite{YS86,Ourj07}, are considered as resources.
This reduces $N$ by many orders of magnitude compared to OCQC, but at the cost of a lower $\eta_{\rm th}$. As seen form Tab.~\ref{tab:para}, the $\eta_{\rm th}$ is smaller by an order of magnitude than OCQC. 

A hybrid-qubit-based QC (HQQC)~\cite{LJ13} scheme  uses optical hybrid states instead of coherent superposition states. HQQC  offers a better value of  $\eta_{\rm th}$ and resource scaling than CSQC.  If different kind of hybrid qubits~\cite{KLJ16} are employed for telecorrection, one would speculate a better resilience against photon loss in HQQC.
In linear optical FTQC, the recent HTQC~\cite{OTJ19} offers the best $\eta_{\rm th}$ and resource efficiency known to date. However,  PHTQC-2 and PHTQC-3 lead to even better $\eta_{\rm th}$ than  HTQC at the cost of incurring slightly more resources. Nevertheless, the new schemes remain resource efficient in marked contrast to all other linear optical schemes for FTQC.

We stress caution by noting  that in OCQC, PLOQC, EDQC and TPQC, the two noise parameters, $\eta$ and the  computational error rate, are independent. However, these parameters are  interdependent in HTQC, PHTQC, CSQC, HQQC and MBQC. Moreover, in the former schemes, the computational error is depolarizing in nature whereas in the latter schemes, it is a result of dephasing caused by photon loss.  It is important to note $\eta_{\rm th}$'s claimed by OCQC,  PLOQC, EDQC and TPQC are valid only for zero computational errors, which is unrealistic since photon losses typically cause computational errors.


\section{Discussion and Conclusion}
\label{sec:conclusion}
  
In pushing hybrid qubit quantum computing to the limit, we establish  postselection schemes for the creation of star cluster states  and utilize multiple hybrid Bell-state measurements per edge creation to build a Raussendorf lattice for fault-tolerant quantum computation. Compared to a recently published hybrid qubit scheme~\cite{OTJ19}, we show that our current  hybrid scheme with postselection can achieve an even higher photon-loss threshold. In particular, we achieve the threshold values of $5\times10^{-3}$ and $5.7\times10^{-3}$ with two respective subvariants of postselection scheme introduced in this work. They represent an approximately 50\% improvement compared to the previous  scheme without postselection ($3.3\times10^{-3}$)~\cite{OTJ19}.

This enhancement comes from the desirable fact that the hybrid-qubit scheme with postselection can have a high success rate of entangling operations without the need to use hybrid qubits of large coherent amplitudes. Consequently, the current scheme benefits from weaker dephasing effects arising from photon loss. We also show that a larger photon loss threshold comes at a nominal increase in resource overhead of about one order of magnitude in comparison with that for the hybrid qubit scheme without postselection. In terms of hardware design, this additionally requires switching circuits to support postselection and multiple hybrid Bell-state measurements. Therefore, the ballistic character of the previous hybrid qubit scheme is sacrificed in exchange for higher photon-loss tolerance. 

From these findings, we now confirm that all hybrid-qubit schemes permit significantly higher operational photon loss and computational error rates, by an order of magnitude compared to other optical schemes~\cite{DHN06,Cho07,HHGR10,LPRJ15,HFJR10,LRH08,LJ13}. Although the  optical-cluster scheme~\cite{DHN06}
provides a slightly larger photon loss threshold compared to the previous ballistic hybrid-qubit scheme~\cite{OTJ19}, we have shown that our current scheme can provide an even larger threshold values. We also demonstrate an overall superiority in resource efficiency of our current scheme. If the failure rate of hybrid Bell-state measurements is large, postselection of higher intensity is required and would render its performance comparable to discrete variable schemes. Eventually, using hybrid qubits offers no resource advantage over the discrete variable scheme in Ref~\cite{LHMB15}.

On hindsight, since using smaller coherent amplitudes in postselection hybrid schemes boosts photon-loss thresholds, it naturally supports the logic that a Raussendorf lattice built with only discrete-variable  qubits could offer an even higher photon loss threshold, albeit at higher resource costs. 

Proposals to generate optical hybrid states, without cross-Kerr non-linearity, using only linear optical elements and photon detectors were made in Refs.~\cite{KJ15, LYHW18, Huang19, Sergey19}. 
Sophisticated manipulations of time-bin and wave-like degrees of freedom have also opened interesting routes to generating such entangled states~\cite{JZK+14, Morin14, Alexander17, SUT+18, Cavailles18,Guccione20, GBTD20}.
These achievements pave the way to practical hybrid qubit quantum computing.



{\it Acknowledgments.---} We thank  Austin G. Fowler for useful discussions and suggestions. 
This work was supported by National Research Foundation of Korea (NRF) grants funded by the Korea government (Grants No.~2019M3E4A1080074 and No.~2020R1A2C1008609).
Y.S.T. was supported by an NRF grant funded by the Korea government (Grant No. NRF-2019R1A6A1A10073437). 
S.W.L. acknowledges support from the National Research Foundation of Korea (2020M3E4A1079939) and the KIST institutional program (2E30620).

%


\end{document}